\documentclass[iop]{emulateapj}
\usepackage{appendix}

\begin{document}

\newcommand{\mirlum}{L_{\rm 8}}
\newcommand{\ebmv}{E(B-V)}
\newcommand{\lha}{L(H\alpha)}
\newcommand{\lir}{L_{\rm IR}}
\newcommand{\lbol}{L_{\rm bol}}
\newcommand{\luv}{L_{\rm UV}}
\newcommand{\rs}{{\cal R}}
\newcommand{\ugr}{U_{\rm n}G\rs}
\newcommand{\ks}{K_{\rm s}}
\newcommand{\gmr}{G-\rs}
\newcommand{\hi}{\text{\ion{H}{1}}}
\newcommand{\nhi}{N(\text{\ion{H}{1}})}
\newcommand{\lognhi}{\log[\nhi/{\rm cm}^{-2}]}
\newcommand{\molh}{\text{H}_2}
\newcommand{\nmolh}{N(\molh)}
\newcommand{\lognmolh}{\log[\nmolh/{\rm cm}^{-2}]}

\title{SPECTROSCOPIC MEASUREMENTS OF THE FAR-ULTRAVIOLET DUST
  ATTENUATION CURVE AT $z\sim 3$\altaffilmark{*}} \author{\sc Naveen
  A. Reddy\altaffilmark{1,6}, Charles C. Steidel\altaffilmark{2}, Max
  Pettini\altaffilmark{3}, \& Milan Bogosavljevi\'c\altaffilmark{4,5}}

\altaffiltext{*}{Based on data obtained at the W.M. Keck Observatory,
  which is operated as a scientific partnership among the California
  Institute of Technology, the University of California, and NASA, and
  was made possible by the generous financial support of the W.M. Keck
  Foundation.}

\altaffiltext{1}{Department of Physics and Astronomy, University of California, 
Riverside, 900 University Avenue, Riverside, CA 92521, USA}
\altaffiltext{2}{Department of Astronomy, California Institute of Technology, 
MS 105--24, Pasadena,CA 91125, USA}
\altaffiltext{3}{Institute of Astronomy, Madingley Road, Cambridge
CB3 OHA, UK}
\altaffiltext{4}{Astronomical Observatory, Belgrade, Volgina 7, 11060 Belgrade, Serbia}
\altaffiltext{5}{New York University Abu Dhabi, P.O. Box 129188, Abu Dhabi, UAE}
\altaffiltext{6}{Alfred P. Sloan Research Fellow}

\slugcomment{DRAFT: \today}

\begin{abstract}

We present the first measurements of the shape of the far-ultraviolet
(far-UV; $\lambda=950-1500$\,\AA) dust attenuation curve at high
redshift ($z\sim 3$).  Our analysis employs rest-frame UV spectra of
933 galaxies at $z\sim 3$, 121 of which have very deep spectroscopic
observations ($\ga 7$\,hrs) at $\lambda=850-1300$\,\AA, with the Low
Resolution Imaging Spectrograph on the Keck Telescope.  By using an
iterative approach in which we calculate the ratios of composite
spectra in different bins of continuum color excess, $\ebmv$, we
derive a dust curve that implies a lower attenuation in the far-UV for
a given $\ebmv$ than those obtained with standard attenuation curves.
We demonstrate that the UV composite spectra of $z\sim 3$ galaxies can
be modeled well by assuming our new attenuation curve, a high covering
fraction of $\hi$, and absorption from the Lyman-Werner bands of
$\molh$ with a small ($\la 20\%$) covering fraction.  The low covering
fraction of $\molh$ relative to that of the $\hi$ and dust suggests
that most of the dust in the ISM of typical galaxies at $z\sim 3$ is
unrelated to the catalysis of $\molh$, and is associated with other
phases of the ISM (i.e., the ionized and neutral gas).  The far-UV
dust curve implies a factor of $\approx 2$ lower dust attenuation of
Lyman continuum (ionizing) photons relative to those inferred from the
most commonly assumed attenuation curves for $L^{\ast}$ galaxies at
$z\sim 3$.  Our results may be utilized to assess the degree to which
ionizing photons are attenuated in \ion{H}{2} regions or, more
generally, in the ionized or low column density ($\nhi \la
10^{17.2}$\,cm$^{-2}$) neutral ISM of high-redshift galaxies.

\end{abstract}


\section{INTRODUCTION}
\label{sec:intro}

The dust attenuation curve is a principal tool for inferring the
absorption and scattering properties of interstellar dust grains, the
spatial distribution of that dust relative to stars, and the intrinsic
stellar populations of galaxies.  Observations of nearby galaxies have
yielded detailed information on both the normalization and shape of
the dust curve (e.g., \citealt{calzetti94, calzetti00, johnson07a,
  wild11}).  In contrast, our knowledge of the dust curve (and how the
curve varies) at high redshift has only recently improved with large
multi-wavelength and spectroscopic surveys (e.g., \citealt{noll09,
  buat11, buat12, kriek13, scoville15, reddy15, zeimann15, salmon15}).

Despite these advances, the shape of the attenuation curve at
far-ultraviolet (far-UV) wavelengths, in particular at $\lambda \la
1250$\,\AA, is unconstrained at high redshift.  Establishing the shape
of this curve at these wavelengths is imperative for several reasons.
First, dust obscuration of starlight reaches its maximum in the
far-UV, such that even small differences in the continuum color
excess, $\ebmv$, may lead to large variations in the inferred
attenuation of far-UV light.  Hence, inferences of the intrinsic
formation rate of the very massive stars that dominate the far-UV
requires knowledge of the shape of the dust curve at these
wavelengths.  Second, the translation between reddening and far-UV
color depends on the shape of the far-UV attenuation curve.
Therefore, different shapes of the far-UV curve will result in
different color criteria to select galaxies at intermediate redshifts
($1.5\la z\la 4.0$) where the Ly$\alpha$ forest is sufficiently thin
and where the selection may include photometric bands that lie
blueward of Ly$\alpha$.  Stated another way, fixed far-UV color
selection criteria will select galaxies at slightly different
redshifts depending on the shape of the UV dust attenuation curve, and
may affect the inferred selection volumes that are necessary for
computing UV luminosity functions (e.g., \citealt{steidel93,
  steidel03, adelberger04, sawicki06, reddy08}).  Third, the utility
of the Ly$\alpha$ emission line as a probe of the physical conditions
(e.g., dust/gas distribution, gas kinematics) in high-redshift
galaxies requires some knowledge of dust grain properties and the
distribution of the dust in the ISM (e.g., \citealt{meier81,
  calzetti92, charlot93, giavalisco96, verhamme08, scarlata09,
  finkelstein11, hayes11, atek14}), both of which are reflected in the
shape of the attenuation curve.  Fourth, there has been renewed
attention in the modeling of massive stars \citep{eldridge09, brott11,
  levesque12, leitherer14} given evidence of rest-optical line ratios
\citep{steidel14, shapley15} and UV spectral features of $z\sim 2-3$
galaxies (Steidel et~al., in prep) that cannot be modeled
simultaneously with standard stellar population synthesis models.  The
modeling of the far-UV continuum of high-redshift galaxies is
sensitive to the assumed dust attenuation curve.  Lastly, knowledge of
the attenuation curve at or near the Lyman break is needed to deduce
the effect of dust obscuration on the depletion of hydrogen-ionizing
photons, a potentially important ingredient in assessing the
contribution of galaxies to reionization and translating recombination
line luminosities to star-formation rates.

Given the relevance of the far-UV shape of the dust attenuation curve,
we sought to use the existing large spectroscopic samples of Lyman
Break galaxies at $z\sim 3$ to provide the first direct constraints on
the shape of this curve at high redshift.  In
Section~\ref{sec:sample}, we discuss the sample of galaxies and the
procedure to construct far-UV spectral composites.  The calculation of
the far-UV dust attenuation curve is presented in
Section~\ref{sec:faruvcurve}.  In Section~\ref{sec:specfitting}, we
assess whether the dust curve can be used in conjuction with spectral
models to reproduce the composite UV spectra.  The implications of our
results are discussed in Section~\ref{sec:discussion}, and we
summarize in Section~\ref{sec:summary}.  All wavelengths are in
vacuum.  All magnitudes are expressed in the AB system.  We adopt a
cosmology with $H_{0}=70$\,km\,s$^{-1}$\,Mpc$^{-1}$,
$\Omega_{\Lambda}=0.7$, and $\Omega_{\rm m}=0.3$.

\section{SAMPLE SELECTION AND REST-FRAME UV SPECTROSCOPY}
\label{sec:sample}

\subsection{Lyman Break Selected Samples at $z\sim 3$}

The galaxies used in our analysis were drawn from the Lyman Break
selected and spectroscopically confirmed sample described in
\citet{steidel03}.  In brief, these galaxies were selected as $z\sim
3$ candidates based on their rest-frame UV ($U_{\rm n}-G$ and $G-{\cal
  R_{\rm s}}$) colors as measured from ground-based imaging of 17
fields.  Of the 2,347 ${\cal R}_{\rm s}\le 25.5$ Lyman Break galaxy
(LBG) candidates, 1,293 were observed spectroscopically with the
original single channel configuration of the Low Resolution Imaging
Spectrograph (LRIS; \citealt{oke95}) on the Keck telescopes, with a
typical integration time of $\simeq 1.5$\,hr.  The spectroscopy was
accomplished using the 300\,lines\,mm$^{-1}$ grating blazed at
5000\,\AA\, giving a typical observed wavelength coverage of
$\lambda_{\rm obs} \approx 4000-7000$\,\AA\, with
2.47\,\AA\,pix$^{-1}$ dispersion.  The spectral resolution, including
the effects of seeing, is $\approx 7.5$\,\AA\, in the observed frame.
Redshifts were determined from either interstellar absorption lines at
$\lambda_{\rm rest}=1200-1700$\,\AA, or from Ly$\alpha$ emission.  In
some cases, the detection of a single emission line with a continuum
break shortward of that line was used to assign a secure redshift.
Our sample includes only those galaxies with secure emission and/or
absorption line redshifts that were vetted by at least two people and
published in \citet{steidel03}.  Further details of the photometry,
candidate selection, followup spectroscopy, and data reduction are
provided in \citet{steidel03}.

Subsequent deep ($\simeq 7-10$\,hrs) spectroscopy was obtained for a
sample of 121 $2.7<z<3.6$ galaxies with existing spectroscopic
redshifts from the \citet{steidel03, steidel04} samples, with the aim
of obtaining very sensitive observations of the Lyman continuum (LyC)
emission blueward of the Lyman limit ($\lambda < 912$\,\AA)---we refer
to the set of objects with deeper spectroscopy as the ``LyC sample''.
This spectroscopy was obtained using the blue channel of the upgraded
double spectrograph mode of LRIS \citep{steidel04}, which presents a
factor of $3-4$ improvement in throughput at $4000$\,\AA\, relative to
the red-side channel.  The new configuration of LRIS includes an
atmospheric dispersion corrector (ADC) that eliminates wavelength
dependent slit losses.  The deep spectroscopic observations were
obtained using the d500 dichroic with the 400\,lines\,mm$^{-1}$ grism
blazed at 3400\,\AA, yielding a (binned) dispersion of
2.18\,\AA\,pix$^{-1}$.  Including the effect of seeing, the typical
observed-frame resolution is $\approx 7$\,\AA\, for the blue channel
spectra.  Redshifts were determined using the same procedure as for
the \citet{steidel03} sample (see above).  The depth of the
spectroscopic observations for the LyC sample allowed us to assign
absorption line redshifts for $\approx 92\%$ of the objects.  The 121
galaxies of the LyC sample excludes all objects ($\approx 5\%$ of the
sample) where the deep spectra indicated a blend between the target
LBG and a foreground galaxy.  Details of the data reduction and
spectroscopy are provided in Steidel et al. (in prep).

We further excluded AGN from the samples based on the presence of
strong UV emission lines (e.g., Ly$\alpha$ and \ion{C}{4}).  The AGN
fraction in the $z\sim 3$ sample (to the magnitude limit of ${\cal
  R}_{\rm s}=25.5$) based on strong UV emission lines is $\approx 3\%$
(e.g., \citealt{steidel03, reddy08}).  We did not have the requisite
multi-wavelength (infrared, X-ray) data for the majority of the fields
to evaluate AGN using other methods.  However, we note that
optically-identified AGN in UV-selected samples comprise $\approx
70\%$ of AGN identified either from X-ray emission, or from a
mid-infrared excess (see \citealt{reddy06b}), and we do not see
evidence of AGN either in the remaining individual spectra or in the
composite spectra.

\begin{figure}
\plotone{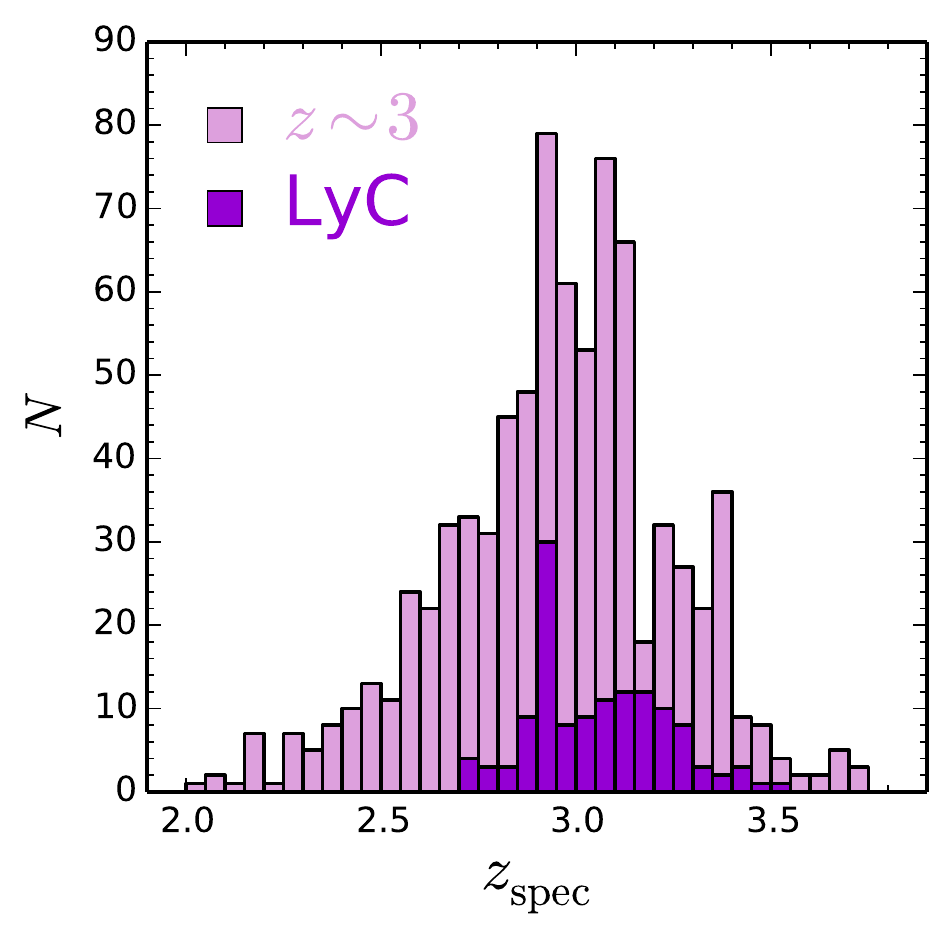}
\caption{Redshift distribution of 933 star-forming galaxies in our
  sample, consisting of objects drawn from \citet{steidel03} ({\em
    light region}) and the deep LyC survey ({\em dark region}).  The
  excess number of redshifts (by a factor of $\ga 3$) in the range
  $2.90<z<2.95$ relative to adjacent redshift bins is due to an
  over-density of galaxies at $z\simeq 2.925$ in the Westphal field.}
\label{fig:zhist}
\end{figure}

Two important advantages of the $z\sim 3$ sample that facilitate
spectroscopy in the far-UV region is the lower sky background in the
observed optical relative to the near-infrared, and the fact that the
Ly$\alpha$ forest is still relatively thin at these redshifts,
allowing for the transmission of sufficient flux blueward of
Ly$\alpha$.  Figure~\ref{fig:zhist} shows the spectroscopic redshift
distribution of our final sample, which includes 933 galaxies.

\subsection{Composite Spectra}
\label{sec:compspec}

\begin{figure*}
\epsscale{1.19}
\plotone{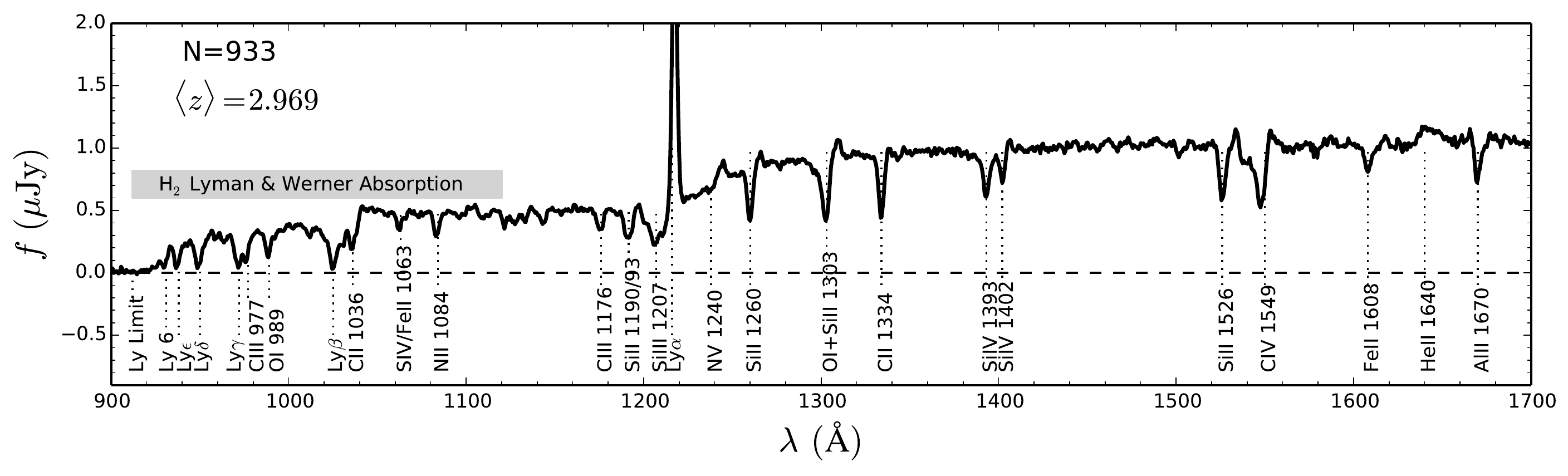}
\caption{Composite spectrum of 933 galaxies at $z\sim 3$, with several
  of the most prominent stellar photospheric and interstellar
  absorption lines indicated.  The far-UV range
  $\lambda=912-1120$\,\AA\, is dominated by unresolved absorption from
  the Lyman and Werner bands of molecular hydrogen ($\molh$).  The
  composite has been normalized by the median flux density in the
  range $\lambda=1400-1500$\,\AA.}
\label{fig:compall}
\end{figure*}

Throughout our analysis, we employed composite spectra which were
computed by combining the individual spectra of galaxies.  We first
made separate composites of the galaxies in the $z\sim 3$ LBG and LyC
samples, and then combined the two resulting composites to obtain a
final composite.  This procedure allows us to take advantage of the
greater spectroscopic depth of the LyC objects in the far-UV range
$\lambda_{\rm rest}=912-1150$\,\AA.

For each galaxy in the $z\sim 3$ LBG sample, we determined the
redshift using either emission (e.g., Ly$\alpha$) and/or interstellar
absorption lines.  As these features are typically either redshifted
(in the case of emission) or blueshifted (in the case of absorption)
due to bulk flows of gas, we computed a systemic redshift from the
emission and/or absorption line redshifts using the relations in
\citet{steidel10}.  We then shifted each galaxy's spectrum into the
rest frame using the systemic redshift, and interpolated the spectrum
to a grid with a wavelength spacing of $\delta\lambda = 0.5$\,\AA.
The error spectrum for each galaxy was shifted and interpolated
accordingly, and then multiplied by $\sqrt{2}$ to account for
wavelength resampling from the original dispersion of the reduced
spectra of 1\,\AA\,pix$^{-1}$.  Each spectrum was scaled according to
the median flux density in the range $\lambda=1400-1500$\,\AA.  The
individual galaxy spectra were then averaged, rejecting the upper and
lower three flux density values in each $\delta\lambda$ interval.

The LyC composite was constructed using a similar procedure, with two
exceptions.  First, we did not scale the LyC spectra as the
spectroscopic setup (with the d500 dichroic) precluded coverage of
$\lambda=1400-1500$\,\AA\, within the blue channel of LRIS.  Second,
we combined the LyC spectra using a weighted average, where the
weights are $w(\lambda) = 1/\sigma^2(\lambda)$, where
$\sigma(\lambda)$ is the error spectrum.  We still rejected the upper
and lower three weighted flux density values in each $\delta\lambda$
interval.

The LyC composite was scaled to have the same median flux density in
the range $\lambda=1100-1150$\,\AA\, as that of the $z\sim 3$ LBG
composite.  The final composite results from combining the scaled LyC
composite at $\lambda\le 1150$\,\AA\, and the $z\sim 3$ LBG composite
at $\lambda>1150$\,\AA.  We confirmed that this combination does not
introduce a systematic bias in the level of the stellar continuum of
the composite below and above $\lambda=1150$\,\AA.  The mean redshifts
of the objects contributing to the composite spectrum at the blue
($912$\,\AA) and red ($1600$\,\AA) ends of the composite spectrum are
$\langle z\rangle = 3.058$ and $2.940$, respectively.  Given this
small difference in redshift, corresponding to $\delta t \approx
92$\,Myr, we assumed that the LBG spectral properties do not vary over
this short time interval.  The final composite is shown in
Figure~\ref{fig:compall}, normalized such that the median flux density
in the range $\lambda=1400-1500$\,\AA\, is unity.  The error spectrum
corresponding to the composite was computed as the weighted
combination of the error spectra of the individual objects
contributing to the composite.  The error in the mean flux of the
composite is $<0.1$\,$\mu$Jy for the scaling and full wavelength range
shown in Figure~\ref{fig:compall}.

\section{FAR-UV ATTENUATION CURVE}
\label{sec:faruvcurve}

The method of computing the far-UV attenuation curve, $k(\lambda)$, is
based on taking the ratios of composite spectra in bins of $\ebmv$.
The attenuation curve is defined as
\begin{equation}
k(\lambda) \equiv \frac{-2.5}{\ebmv_{i+1}-\ebmv_{i}}\log\left[\frac{f_{i+1}(\lambda)}{f_{i}(\lambda)}\right],
\label{eq:attcurve}
\end{equation}
where $f_i(\lambda)$ and $\ebmv_i$ refer to the average flux density and
average $\ebmv$, respectively, for objects in the {\em i}th bin of $\ebmv$.

Our procedure involves computing the $\ebmv$ for each galaxy based on
its rest-UV ($\gmr$) color, and an assumed intrinsic template and
attenuation curve.  The template is a combination of the \citet{rix04}
models and the 2010 version of Starburst99 spectral synthesis models
in the far-UV \citep{leitherer99, leitherer10}.  The models were
computed assuming a stellar metallicity of $0.28Z_\odot$, based on
current solar abundances \citep{asplund09}, with a constant
star-formation history and age of 100\,Myr.  This particular
combination of parameters was found to best reproduce the UV stellar
wind lines blueward of Ly$\alpha$ (e.g., \ion{C}{3}\,$\lambda$1176),
as well as the \ion{N}{5}\,$\lambda$1240 and \ion{C}{4}\,$\lambda$1549
P-Cygni profiles.  The star-formation history and age of the model
($100$\,Myr) were chosen to be consistent with the results from
stellar population modeling of $z\sim 3$ LBGs presented in
\citet{shapley01} and \citet{reddy12b}.  On average, such galaxies
exhibit UV-to-near-IR SEDs and bolometric SFRs that are well-described
by either a rising or constant star-formation history
\citep{reddy12b}.  For these star-formation histories, the UV spectral
shape remains constant for ages $\ga 100$\,Myr as the relative
contribution to the UV light from O and B stars stabilizes after this
point.  Moreoever, there are no significant correlations between
reddening and either star-formation history or age for galaxies in our
sample \citep{reddy12b}.  Thus, adopting a different model
star-formation history and age will serve to systematically shift the
computed $\ebmv$, but the value of the attenuation curve, which
depends on the {\em difference} in the average $\ebmv$ from bin-to-bin
(Equation~\ref{eq:attcurve}), will be unaffected.  Nebular continuum
emission was added to the stellar template to produce a final model
that we subsquently call the ``intrinsic template''.\footnote{The
  model that includes nebular continuum emission results in $\ebmv$
  that are systematically smaller (because the spectrum of the
  template is redder), typically by $\delta E(B-V) \simeq 0.03-0.05$,
  than a model that includes only the stellar light.}

\begin{figure*}
\epsscale{1.00}
\plotone{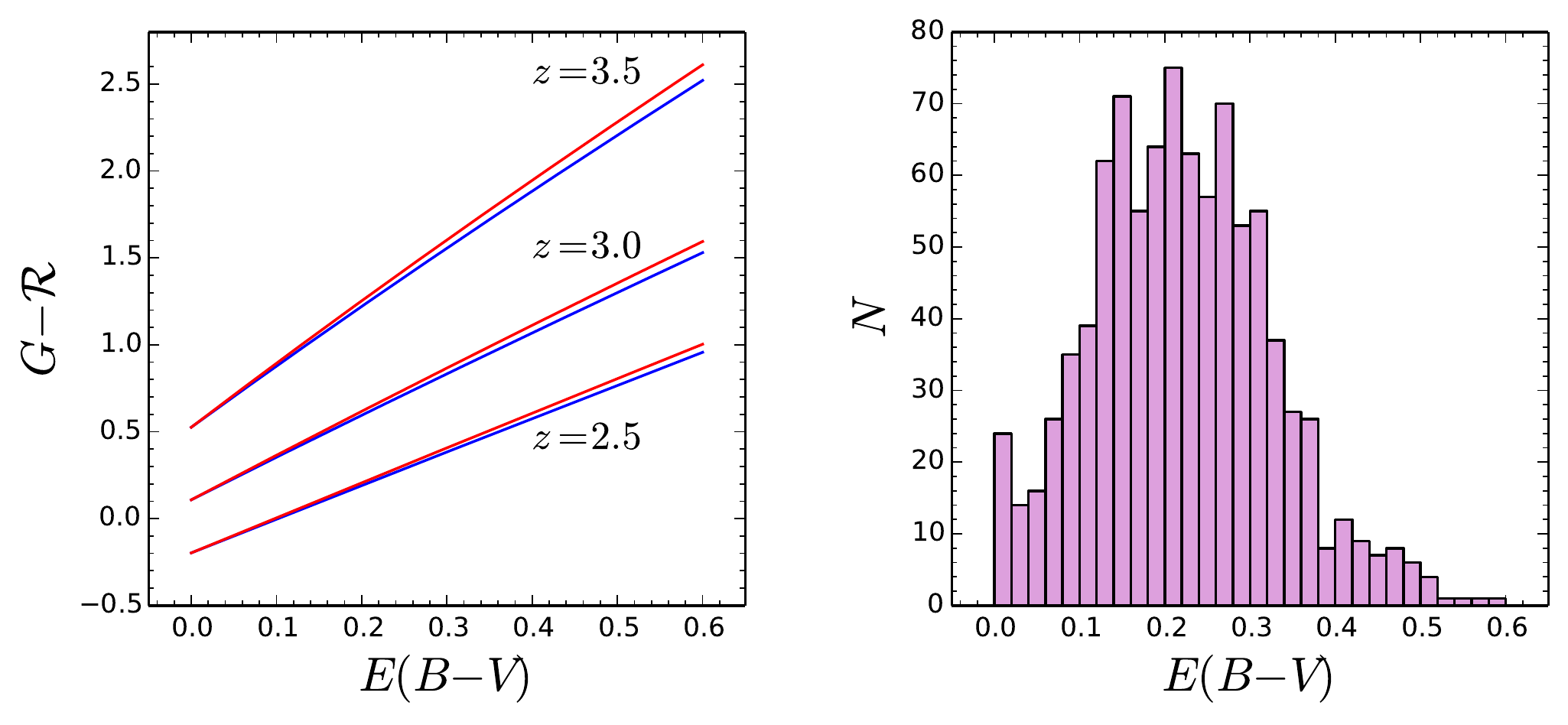}
\caption{{\em Left:} Relationship between $\ebmv$ and $\gmr$ color
  using the intrinsic template and an IGM opacity based on the $\nhi$
  column density distribution of \citet{rudie13}.  The relationship is
  shown for three redshifts, assuming either the \citet{calzetti00} or
  \citet{reddy15} attenuation curves, indicated by the {\em red} and
        {\em blue} lines, respectively.  {\em Right:} Distribution of
        $\ebmv$ for the 933 galaxies in our sample, where the $\ebmv$
        have been computed assuming the \citet{calzetti00} attenuation
        curve.}
\label{fig:colordist}
\end{figure*}

We generated a grid of $\gmr$ colors by reddening the intrinsic
template assuming $\ebmv=0.00-0.60$, with increments of $\delta\ebmv =
0.01$, and two different attenuation curves: the \citet{calzetti00}
and \citet{reddy15} curves.  The former curve is the most commonly
used attenuation curve for high-redshift galaxies.  The latter curve
is our most recent determination using a sample of $224$ star-forming
galaxies at redshifts $z=1.36-2.59$ from the MOSFIRE Deep Evolution
Field (MOSDEF; \citealt{kriek15}) survey.  While lower in redshift
than most of the galaxies examined here, the MOSDEF sample covers
roughly the same range in stellar mass and star-formation rate as the
latter.  The reddened templates were redshifted and attenuated by the
average IGM opacity appropriate for that redshift.  The average
opacity was determined by randomly drawing line-of-sight absorber
column densities from the distribution measured in the Keck Baryonic
Structure Survey (KBSS; \citealt{rudie13}), and then taking an average
over many lines-of-sight.  As the attenuation curve depends only on
the ratio of the composite spectra and the difference in their
$\ebmv$, adopting a different prescription for the IGM opacity will
not affect the shape of the attenuation curve.  The models were then
multiplied by the {\em G}- and $\rs$-band transmission filters to
produce a model-generated $\gmr$ color associated with each $\ebmv$.
The relationship between $\gmr$ and $\ebmv$, and the distribution of
$\ebmv$ for the 933 galaxies in our sample, are shown in
Figure~\ref{fig:colordist}.

\begin{figure*}
\epsscale{1.15}
\plotone{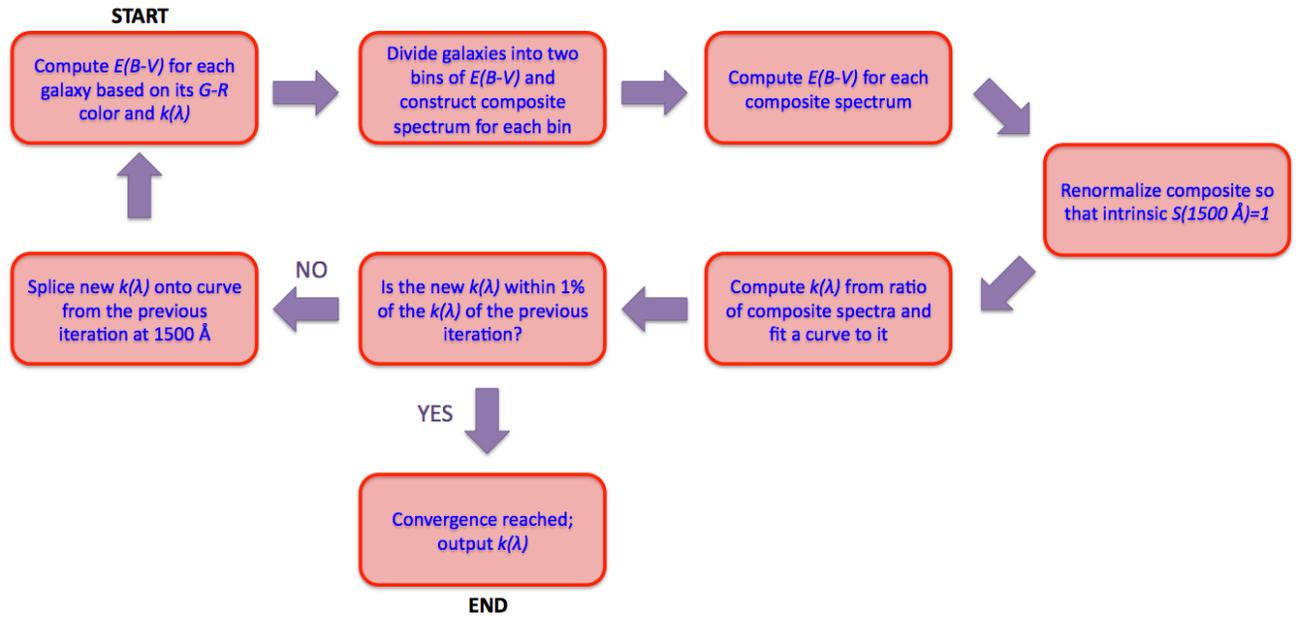}
\caption{Flowchart of the iterative procedure used to compute the
  attenuation curve, $k(\lambda)$.  See the text for a detailed
  description of each step in the derivation.}
\label{fig:flowchart}
\end{figure*}

We developed an iterative derivation of the attenuation curve which
has 7 main steps: (1) compute the $\ebmv$ of each object in the
sample; (2) bin the objects by $\ebmv$ and construct the composite for
each bin; (3) determine the average $\ebmv$ that best fits the
composite; (4) renormalize the composite to be referenced to a
dust-corrected spectrum with a $1500$\,\AA\, flux density equal to
unity; (5) compute the $k(\lambda)$ curve from the ratio of the
composite spectra in different $\ebmv$ bins, and fit a curve to it;
(6) compute the variation between the new $k(\lambda)$ and that of the
previous iteration; and (7) repeat steps (1) through (6) until the
shape of the curve converges.  A flowchart of the procedure is
depicted in Figure~\ref{fig:flowchart}.

Our calculation presents several advantages.  First, performing this
analysis using the composite spectra of many galaxies allows us to
average over stochastic variations in the Ly$\alpha$ forest so that
the resulting composites can be used reliably to measure the far-UV
shape of the dust attenuation curve.  Second, as noted above, the
average stellar populations of galaxies contributing to each bin of
$\ebmv$ are similar.  As the far-UV portion of the spectrum is
dominated by only the most massive stars, the shape of the spectrum
remains constant after $\approx 100$\,Myr and is insensitive to the
presence of older stellar populations.  Third, the stellar mass
distribution of $z\sim 3$ LBGs in our sample, which spans the range
from a few $\times 10^9$\,$M_\odot$ to $\approx 10^{11}$\,$M_\odot$,
implies an $\approx 0.3$\,dex difference in mean oxygen abundance from
the lowest to highest stellar mass galaxies in our sample, based on
the stellar mass-metallicity relation at $z\sim 3$ \citep{troncoso14}.
Assuming a similar range of stellar metallicities for a fixed
star-formation history and IMF implies a $\la 10\%$ difference in the
intrinsic UV slope across the range of galaxies considered here.
Thus, variations in the composite spectral shapes are driven primarily
by changes in dust attenuation, rather than by differences in stellar
metallicity.  The details of our dust attenuation curve calculation
are presented below.

\subsection{Derivation}
\label{sec:derivation}

In step 1, the observed $\gmr$ color of each galaxy was corrected for
Ly$\alpha$ emission or absorption as measured from the rest-UV
spectrum.  This line-corrected color was compared to the
model-generated $\gmr$ colors for each attenuation curve in order to
find the appropriate $\ebmv$.  In step 2, the galaxies were sorted and
divided into two bins of $\ebmv$, a ``low'' (``blue'') and ``high''
(``red'') $\ebmv$ bin, with roughly an equal number of galaxies in
each bin (466 and 467 objects, respectively).  The composite spectrum
was computed for the objects in each bin of $\ebmv$, resulting in a
``blue'' and ``red'' $\ebmv$ composite, following the procedure
specified in Section~\ref{sec:compspec}.

For step 3, the best-fit $\ebmv$ for each composite spectrum was
determined by taking the same intrinsic template specified above and
reddening it according to the \citet{calzetti00} and \citet{reddy15}
attenuation curves using the same range of $\ebmv$ specified above.
These models were then redshifted to the mean redshift of objects
contributing to the composite, and attenuated according to the IGM
opacity appropriate for that redshift.  The model that best fits the
composite spectrum then determines the $\ebmv$.  We considered the
largely line-free wavelength windows listed under ``Composite
$\ebmv$'' in Table~\ref{tab:waves} when comparing the model spectrum
to the composite.  The best-fit $\ebmv$ was found by $\chi^2$
minimization of the model with respect to the composite spectrum.  The
$\ebmv$ computed in this way agreed well with the value inferred from
the average $\gmr$ color of objects contributing to the composite.

\begin{deluxetable}{lc}
\tabletypesize{\footnotesize}
\tablewidth{0pc}
\tablecaption{Wavelength Windows for Fitting}
\tablehead{
\colhead{Type of Fitting} &
\colhead{$\lambda$\,(\AA)}}
\startdata
Composite $\ebmv$\tablenotemark{a} & 1268-1293 \\
 & 1311-1330 \\
 & 1337-1385 \\
 & 1406-1521 \\
 & 1535-1543 \\
 & 1556-1602 \\
 & 1617-1633 \\
\\
$k(\lambda)$\tablenotemark{b} & 950-970 \\
 & 980-985 \\
 & 994-1017 \\
 & 1040-1182 \\
 & 1268-1500 \\
\\
$\nhi$\tablenotemark{c} & 967-974 \\
 & 1021-1029 \\
 & 1200-1203 \\
 & 1222-1248 \\
\\
$\nmolh$\tablenotemark{d} & 941-944 \\
 & 956-960 \\
 & 979-986 \\
 & 992-1020
\enddata
\tablenotetext{a}{Windows used to determine the best-fit $\ebmv$
corresponding to a composite spectrum.}
\tablenotetext{b}{Windows used to determine the best polynomial
fit to the far-UV shape of the attenuation curve, $k(\lambda)$.}
\tablenotetext{c}{Windows used to determine the best-fit column density
$\nhi$ and covering fraction $f_{\rm cov}(\hi)$ of neutral hydrogen.}
\tablenotetext{d}{Windows used to determine the best-fit column density
$\nmolh$ and covering fraction $f_{\rm cov}(\molh)$ of molecular hydrogen.}
\label{tab:waves}
\end{deluxetable}

\begin{figure*}
\epsscale{1.00}
\plotone{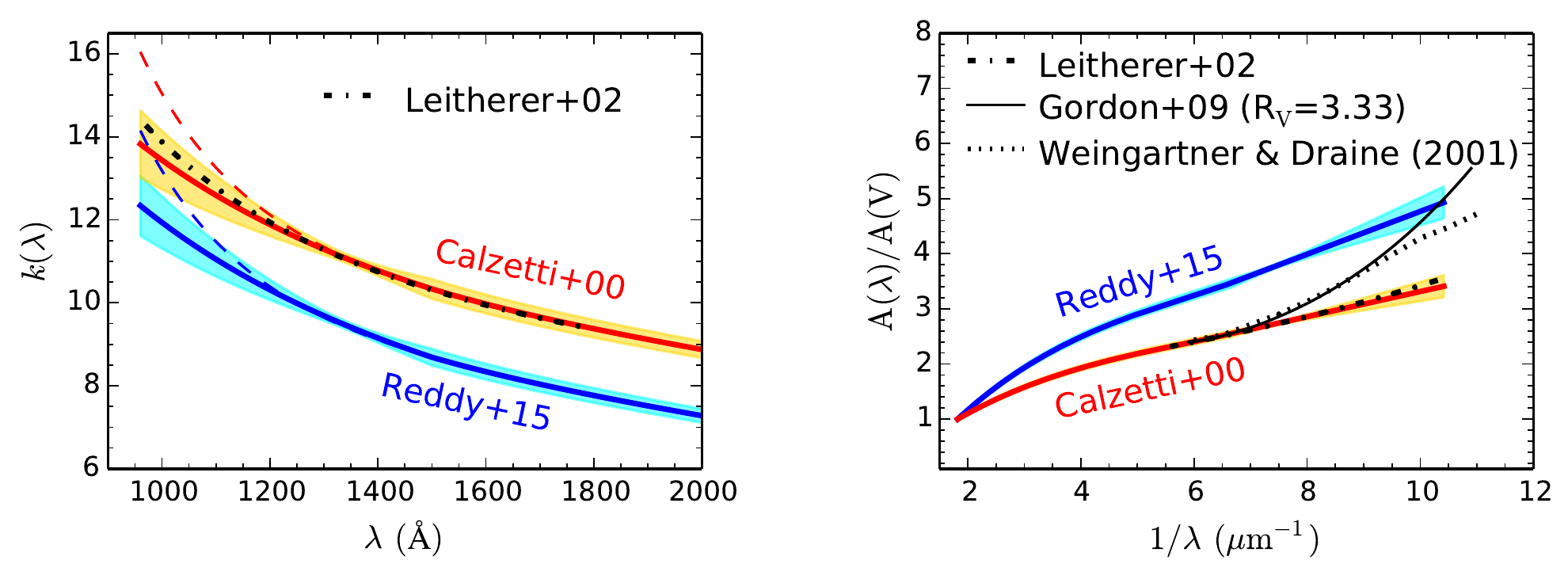}
\caption{{\em Left:} shapes of the far-UV attenuation curves derived
  in this work, normalized to the \citet{reddy15} and
  \citet{calzetti00} attenuation curves at $\lambda=1500$\,\AA\, ({\em
    blue} and {\em red} solid lines, respectively).  The dashed lines
  indicate polynomial extrapolations of the \citet{reddy15} and
  \citet{calzetti00} attenuation curves based on the shapes of these
  curves at $\lambda \ga 1200$\,\AA.  Also shown is the far-UV
  attenuation curve for local galaxies from \citet{leitherer02}.  {\em
    Right:} $A(\lambda)/A({\rm V})$ versus $1/\lambda$ for the same
  curves shown in the left panel, where $A(\lambda)$ is the
  attenuation in magnitudes at wavelength $\lambda$.  The average
  Galactic extinction curve from \citet{gordon09}, and the model
  extinction curve from \citet{weingartner01} are shown for
  comparison.  The shaded regions in both panels indicate the total
  random error in our calculation of the far-UV attenuation curve.}
\label{fig:attcurve}
\end{figure*}

In step 4, the composite was divided by the factor $f(1500)
10^{0.4\ebmv k(1500)}$, where $f(1500)$ and $k(1500)$ are the flux
density and attenuation curve, respectively, at $\lambda=1500$\,\AA.
This renormalization forces the dust-corrected spectrum (i.e., the
spectrum obtained when de-reddening the composite according to its
$\ebmv$ for a given attenuation curve) to have $f_{1500}^{\rm dered} =
1$.  The new shape of the attenuation curve is computed in step 5
using Equation~\ref{eq:attcurve}.  We fit the new $k(\lambda)$ with a
polynomial of the form $p(\lambda) = a_0 + a_1/\lambda$, where
$\lambda$ is the wavelength in $\mu$m.  In fitting $k(\lambda)$, we
only considered windows in the far-UV ($\lambda<1500$\,\AA) that are
free of $\hi$ lines, and which are listed in Table~\ref{tab:waves}.
While the windows do include some metal lines, we confirmed by
excluding the wavelength regions corresponding to the weaker
interstellar and stellar absorption lines that the presence of such
lines did not significantly alter the fit to the attenuation curve.
In addition, we used the errors in the attenuation curve (see below)
to determine the best-fit polynomial using $\chi^2$ minimization.

In step 6, we determined if the new $k(\lambda)$ computed in step 5
differed from the assumed $k(\lambda)$ (i.e., the one from the
previous iteration) by more than $1\%$ at any given wavelength.  If
so, the new $k(\lambda)$ curve blueward of $1500$\,\AA\, was spliced
onto the $k(\lambda)$ from the previous iteration (note that the curve
redward of $1500$\,\AA\, remains fixed throughout the iterative
procedure).  This new version of $k(\lambda)$ was then used to repeat
steps 1 through 6.  Convergence in the shape of the attenuation
curve---defined such that the curve varies by no more than $1\%$ at
any wavelength from that of the previous iteration---was reached
within just two or three iterations.  The curves obtained when
initially assuming the \citet{reddy15} and \citet{calzetti00}
attenuation curves are shown in Figure~\ref{fig:attcurve}.  The
best-fit polynomial to the far-UV curve with the \citet{calzetti00}
normalization at $1500$\,\AA, and which takes into account the random
errors discussed below, is
\begin{eqnarray}
k_{\rm Calz}(\lambda) = (4.126\pm 0.345) + (0.931\pm 0.046)/\lambda, \nonumber \\
0.095\le \lambda \le 0.150\,\mu{\rm m}.
\label{eq:kcalzfit}
\end{eqnarray}
As shown in Figure~\ref{fig:attcurve}, this polynomial is essentially
identical to the fit presented in \citet{calzetti00} for $1250\le
\lambda\le 1500$\,\AA, but deviates from the former curve at shorter
wavelengths ($\lambda < 1250$\,\AA).  Similarly, the functional fit to
the curve with the \citet{reddy15} normalization at $1500$\,\AA\, is
\begin{eqnarray}
k_{\rm Reddy}(\lambda) = (2.191 \pm
0.300) + (0.974 \pm 0.040)/\lambda,
\end{eqnarray}
defined over the same wavelength range specified in
Equation~\ref{eq:kcalzfit}.

\subsection{Random and Systematic Uncertainties}

There are several sources of random error in our calculation of the
far-UV attenuation curve.  First are the measurement errors in the
composite spectra (Section~\ref{sec:compspec}).  Second are
measurement uncertainties in the $\gmr$ colors, which can scatter
galaxies from one $\ebmv$ bin to another.  Third are deviations away
from the attenuation curve from one galaxy to another (e.g., due to
differences in dust/stars geometries), which will affect the $\ebmv$
inferred for a given $\gmr$ color.  Fourth are random variations in
the IGM opacity when averaged over the 466 and 467 sightlines (70 and
60 sightlines for $\lambda<1150$\,\AA) that contribute to the two
$\ebmv$ bins.  The combined effect of these sources of random error
was quantified by calculating $k(\lambda)$ many times using the same
procedure described above, where each time we (a) varied the $\gmr$
color of each galaxy according to its measurement uncertainty; (b)
perturbed the $\ebmv$ associated with the $\gmr$ color for a given
attenuation curve by $\Delta\ebmv = 0.1$ based on the intrinsic
scatter of galaxies used to calibrate the \citet{meurer99} and
\citet{calzetti00} relations, but ensuring that $\ebmv \ge 0$; (c)
randomly drew 466 and 467 sightlines (and 70 and 60 sightlines) based
on the \ion{H}{1} column density distribution measured from KBSS; and
(d) perturbed the composite spectra according to the corresponding
error spectra.  We then computed the dispersion in the fits of the
attenuation curve obtained at each wavelength from these simulations
to estimate the random error in the attenuation curve.  The total
random error obtained in this way is indicated by the shaded regions
around the attenuation curves shown in Figure~\ref{fig:attcurve}.
Note that while the random error increases towards bluer wavelengths
at $\lambda \la 1200$\,\AA, this error is still a factor $\ga 3$
smaller than the systematic differences between our new determinations
of $k(\lambda)$ and the extrapolations of the original
\citet{calzetti00} and \citet{reddy15} curves.  Thus, the differences
between the new far-UV curves derived here and the extrapolations of
the original \citet{calzetti00} and \citet{reddy15} curves at $\lambda
\la 1200$\,\AA\, are statistically significant.

The largest source of systematic error in the far-UV curve is the
attenuation curve initially assumed for the calculation
(Section~\ref{sec:compspec}), as demonstrated in
Figure~\ref{fig:attcurve}.  The far-UV shape of $k(\lambda)$ is
similar regardless of whether we have assumed the \citet{reddy15} or
\citet{calzetti00} curves when initially calculating the $\ebmv$ for
galaxies in our sample, but because of the way in which we calculate
$k(\lambda)$, the normalization is fixed to that of the initially
assumed attenuation curve at 1500\,\AA.  Thus, the normalization of
$k(\lambda)$ is unconstrained in our analysis.

A second source of systematic error arises from the fact that the gas
covering fractions and/or column densities of \ion{H}{1} and $\molh$,
as well as ISM metal absorption, might be expected to increase with
$\ebmv$.  Given that we have computed the attenuation curve in the
far-UV where such lines become prevalent (Figure~\ref{fig:compall}),
we must assess whether differences in the average absorption present
in the composites for the low and high $\ebmv$ bins have a significant
impact on our calculation of $k(\lambda)$---such differences would
cause us to over-estimate the value of $k(\lambda)$.  This issue is
addressed in Section~\ref{sec:specfitting}.

Other sources of systematic uncertainty include our choice of the
intrinsic template and the extrapolation of the \ion{H}{1} column
density distributions measured in KBSS to $z>2.8$.  Both of these have
a negligible effect on the derived attenuation curve as systematic
changes in the intrinsic template or the extrapolation used to model
the Ly$\alpha$ forest at $z>2.8$ will tend to systematically shift the
$\ebmv$ of individual objects either up or down, but the ordering of
the galaxies in $\ebmv$ will be preserved.

\subsection{Comparison to Other Attenuation Curves}

Our calculation marks the first direct measurement of the far-UV
attenuation curve at high redshift using spectroscopic data extending
to the Lyman break at $912$\,\AA.  As can be seen from
Figure~\ref{fig:attcurve}, the dust curve derived here rises less
rapidly at $\lambda\la 1200$\,\AA\, than the polynomial extrapolations
of the \citet{reddy15} and \citet{calzetti00} attenuation curves; the
latter were empirically calibrated for $\lambda\ga 1200$\,\AA.

The far-UV attenuation curve derived here agrees well with that of
\citet{leitherer02} for local star-forming galaxies.  For comparison,
Figure~\ref{fig:attcurve} also includes the \citet{gordon09} average
extinction curve measured over 75 individual Galactic sightlines, and
the extinction curve obtained from the dust grain model of
\citet{weingartner01}.  In general, the far-UV attenuation curve has a
shallower dependence on wavelength than the Galactic extinction curve,
though the comparison is complicated by the fact that the curve
derived here is an {\em attenuation} curve which represents an average
reddening across many sightlines.  Thus, the differences observed
between the various curves in the far-UV may be tied to variations in
the spatial geometry of the dust and stars, as well as to changes in
the dust grain size distribution and scattering and absorption
cross-sections.  The new far-UV measurements of the attenuation curves
derived here are subsequently referred to as the ``updated'' versions
of the \citet{reddy15} and \citet{calzetti00} curves.

\section{SPECTRAL FITTING}
\label{sec:specfitting}

\begin{figure*}
\epsscale{1.15}
\plotone{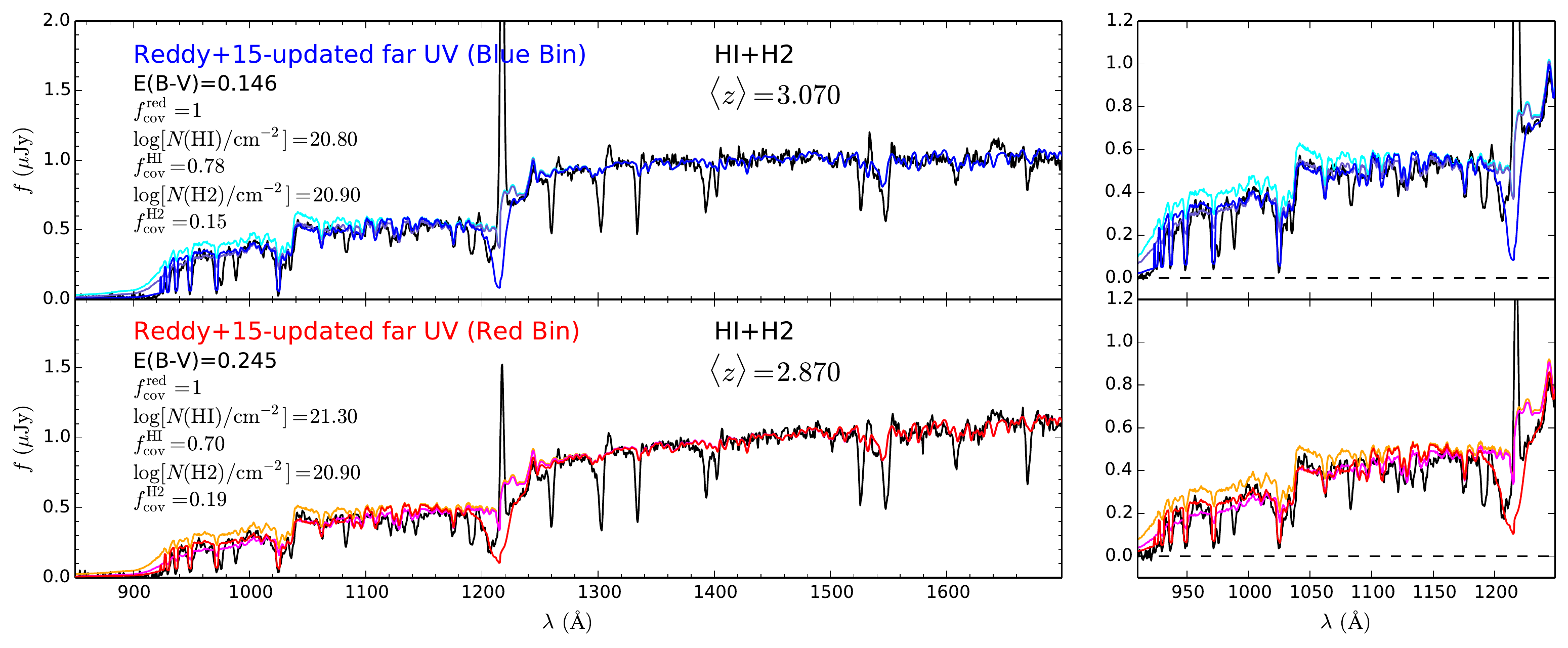}
\caption{Best-fitting models to the composite spectra ({\em black}) in
  the blue ({\em top}) and red ({\em bottom}) bins of $\ebmv$.  The
  intrinsic template was modified by the IGM opacity appropriate for
  the mean redshift of objects contributing to the composites.  The
  fits were then obtained in three steps.  In the first, we found the
  value of $\ebmv$ that minimizes the differences between the model
  and composite spectra in the ``Composite $\ebmv$'' wavelength
  windows specified in Table~\ref{tab:waves}, assuming the updated
  \citet{reddy15} attenuation curve and a unity covering fraction of
  dust $f_{\rm cov}^{\rm red}=1$ (in this first step, $\hi$ and
    $\molh$ absorption are ignored).  These fits are indicated by the
  {\em cyan} and {\em orange} lines in the top and bottom panels,
  respectively.  For comparison, the fits obtained assuming the
  original \citet{reddy15} attenuation curve are shown by the {\em
    dark blue} and {\em magenta} lines in the top and bottom panels,
  respectively; these fits underestimate the continuum at $\lambda <
  1050$\,\AA.  In the second step, we simultaneously varied the column
  density $\nhi$ and covering fraction $f_{\rm cov}(\hi)$ of neutral
  hydrogen until the differences between the model and composite were
  minimized in the ``$\nhi$'' wavelength windows specified in
  Table~\ref{tab:waves}.  In the third and final step, we
  simultaneously varied the column density $\nmolh$ and covering
  fraction $f_{\rm cov}(\molh)$ of molecular hydrogen until the
  differences between the model and composite were minimized in the
  ``$\nmolh$'' wavelength windows specified in Table~\ref{tab:waves}.
  The final fits for the $\ebmv$ and combined $\hi$ and $\molh$
  absorption are indicated by the solid ({\em blue} and {\em red})
  lines in the top and bottom panels, respectively.  For clarity, the
  far-UV region blueward of Ly$\alpha$ is shown in the right panels. }
\label{fig:hih2fit}
\end{figure*}

We assessed whether the UV attenuation curve can adequately describe
the composite spectra of galaxies in our sample and, in the process,
determined the magnitude of any systematic variations in the $\hi$ and
$\molh$ column densities, and metal absorption, between the two bins
of $\ebmv$ used to calculate the curve.  To accomplish this, we
reddened the intrinsic template by various amounts assuming the
updated \citet{reddy15} attenuation curve and a unity covering
fraction of dust ($f_{\rm cov}^{\rm red} = 1$), and determined which
$\ebmv$ results in the best fit to the composite spectrum.  Enforcing
$f_{\rm cov}^{\rm red} = 1$ was motivated by the fact that the
attenuation curve is defined assuming that the entire galaxy is
reddened by some average amount.  The fitting was accomplished using
the ``Composite $\ebmv$'' windows specified in Table~\ref{tab:waves}.
We refer to these as the ``$\ebmv$ Only'' fits (i.e., fits that do not
include the \ion{H}{1} lines or Lyman-Werner absorption from $\molh$),
and they are shown for the two $\ebmv$ bins in
Figure~\ref{fig:hih2fit}.  The uncertainty in $\ebmv$ is
$\sigma(E(B-V))\approx 0.01$ based on perturbing the composite spectra
according to their errors and refitting them many times.  While these
models provide a good fit to the composite spectra at $\lambda
>1260$\,\AA, they clearly overpredict the observed continuum level at
bluer wavelengths, implying that additional absorption must be present
at these wavelengths.

At first glance, the discrepancy between the reddened templates and
the composite spectra in the far-UV may appear to be at odds with the
fact that the attenuation curve was derived from the composite spectra
themselves.  However, recall that the attenuation curve is calculated
from the {\em ratio} of the composite spectra.  Systematic absorption
in the far-UV that is present in the individual composite spectra will
largely ``divide out'' in the calculation of the attenuation curve.

In the next iteration of the spectral fitting, we accounted for
absorption by neutral hydrogen within the galaxies as follows.  We
added $\hi$ Lyman series absorption to the intrinsic template assuming
a Voigt profile with a Doppler parameter $b=125$\,km\,s$^{-1}$ and
varying column densities $\nhi$.  The wings of the line profile are
insensitive to the particular choice of $b$.  We also blueshifted the
$\hi$ absorption lines relative to the model by $300$\,km\,s$^{-1}$ to
provide the best match to the observed wavelengths of the lines in the
composite.  This blueshift is generally attributed to galactic-scale
outflows of gas seen in absorption against the background stellar
continuum.  The composite spectra were compared to the model spectra,
where the latter were computed as:
\begin{eqnarray}
m_{\rm final} & = & 10^{-0.4\ebmv k} \times \nonumber \\
& & [f_{\rm cov}(\hi) m(\hi) + (1-f_{\rm cov}(\hi))m],
\label{eq:hifit}
\end{eqnarray}
where $f_{\rm cov}(\hi)$ is the covering fraction of neutral hydrogen;
$m(\hi)$ and $m$ are the intrinsic templates with and without $\hi$
absorption, respectively; and $\ebmv$ is the value found above (from
the $\ebmv$ Only fitting).  Note that the reddening was applied to the
entire spectrum; i.e., $f_{\rm cov}^{\rm red} = 1$.  These final
models were computed for a range of $\nhi$ and $f_{\rm cov}(\hi)$ and
modified by the IGM opacity appropriate for the average redshift of
each composite.

The final models were compared to the composite spectra in the
``$\nhi$'' wavelength windows specified in Table~\ref{tab:waves} to
find the best-fitting combination of $\nhi$ and $f_{\rm cov}(\hi)$.
The wavelength windows were chosen to avoid regions of strong metal
line absorption (e.g., \ion{C}{3}\,$\lambda$977 and
\ion{C}{2}\,$\lambda$1036) and strong Ly$\alpha$ emission, and include
the core and as much of the wings of Ly$\beta$ and Ly$\gamma$, and the
wings of Ly$\alpha$, particularly the region between Ly$\alpha$ and
\ion{N}{5}\,$\lambda$1238.  The higher Lyman series lines (Ly$\delta$
and greater) were not used due to the larger random errors at these
wavelengths and because these lines begin to overlap.  The covering
fraction is constrained primarily by the depths of the lines, whereas
the column density is most sensitive to the wings.  The fitting
suggests column densities of $\lognhi\approx 20.8-21.3$ and covering
fractions of $f_{\rm cov}(\hi)\approx 0.70-0.78$ for the two bins of
$\ebmv$ (Figure~\ref{fig:hih2fit}).  By perturbing the composite
spectra and re-fitting the data many times, we estimated the
(measurement) errors in $\lognhi$ and $f_{\rm cov}(\hi)$ to be
$\sigma(\lognhi)\approx 0.20$\,dex and $\sigma(f_{\rm cov}(\hi))
\approx 0.01$.  The systematic error in $\lognhi$ from changing the
Doppler parameter by a factor of two from our assumed value is
$\sigma(\lognhi)\approx 0.10$\,dex.

Note that the values of $f_{\rm cov}(\hi)$ derived here are physically
meaningful only if the $\hi$ lines are optically thick {\em and} the
residual flux observed at the $\hi$ line cores is not an artifact of
limited spectral resolution, redshift imprecision, or foreground
contamination.  In the first case, a high $\nhi$ is required to
reproduce the damping wings of Ly$\alpha$, Ly$\beta$, and Ly$\gamma$,
implying optically thick gas.  In the second case, as shown in detail
in Reddy et~al. (2016b), redshift and spectral resolution
uncertainties are not sufficient to account for the residual flux seen
at the $\hi$ line centers.  Briefly, we executed simulations where we
added $\hi$ of varying column densities to the intrinsic spectrum,
replicated many versions of this $\hi$-absorbed spectrum with slightly
different redshifts based on the typical redshift uncertainty for
galaxies in our sample, smoothed these versions with various values of
the spectral resolution, and then produced a composite of all these
versions.  This exercise revealed that we would have had to
systematically underestimate the rest-frame spectral resolution by
more than $50\%$ in order to reproduce the amount of residual flux at
the $\hi$ line centers.  This $50\%$ is much larger than the rms
uncertainties in determining the spectral resolution from the widths
of night sky lines.  Lastly, we ruled out foreground contamination as
a source of the residual flux based on the exclusion of spectroscopic
blends from our sample and the fact that this flux inversely
correlates with reddening, a result that would not be expected if the
flux was dominated by foreground contaminants (see Reddy et~al. 2016b
for more discussion).  Thus, we conclude that the spectra can be
modeled with a partial covering fraction of optically thick $\hi$.

In the final iteration of the fitting, we included $\molh$
Lyman-Werner absorption in the models as follows:
\begin{eqnarray}
m_{\rm final} & = & 10^{-0.4\ebmv k} \times \nonumber \\
& & [(f_{\rm cov}(\hi)-f_{\rm cov}(\molh)) m(\hi) + \nonumber \\
& & f_{\rm cov}(\molh) m(\hi+\molh) + (1-f_{\rm cov}(\hi))m],
\end{eqnarray}
where $f_{\rm cov}(\molh)$ is the covering fraction of molecular
hydrogen, $m(\hi+\molh)$ is the model including both $\hi$ and $\molh$
absorption, and the remaining variables have the same definitions as
in Equation~\ref{eq:hifit}.  This calculation assumes that the $\molh$
is entirely embedded in $\hi$ clouds.

\begin{figure*}
\epsscale{1.15}
\plotone{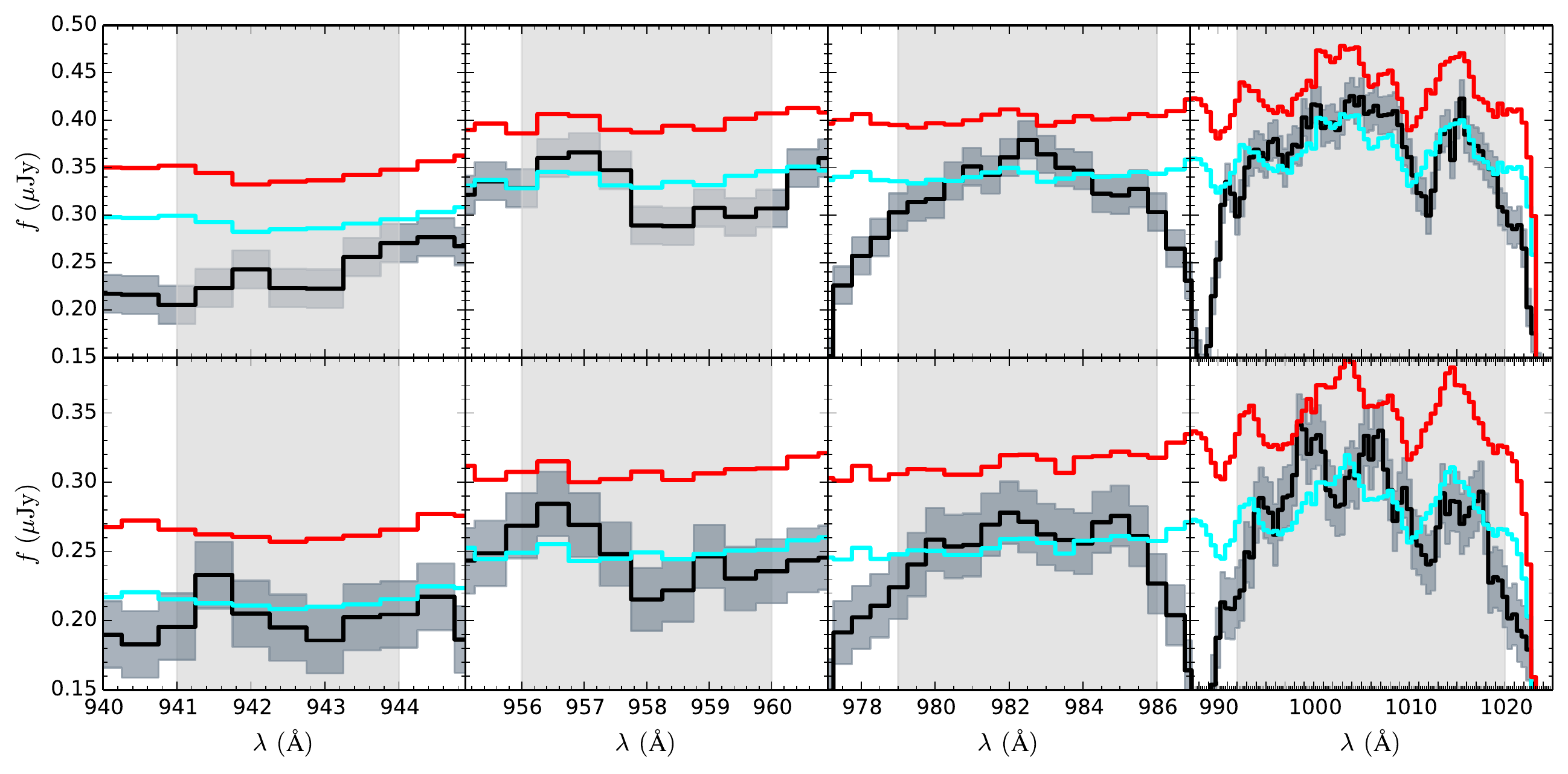}
\caption{Comparison between the composite spectrum ({\em black lines})
  and the best-fit models with ({\em cyan}) and without ({\em red})
  fitting for $\molh$ absorption in the 4 ``$N(\molh)$'' windows
  specified in Table~\ref{tab:waves}.  The wavelength windows are
  indicated by the light shaded regions.  The top and bottom rows show
  the comparison for the composite spectra in the blue and red bins of
  $\ebmv$, respectively.  In general, the fits that include $\molh$
  absorption are in better agreement with the composite spectra than
  the fits that exclude this absorption.  For reference, the $\pm
  1$\,$\sigma$ uncertainty in the composite fluxes are denoted by the
  dark shaded regions.}
\label{fig:h2fit}
\end{figure*}

$\molh$ absorption was modeled using the \citet{mccandliss03}
templates, assuming a Doppler parameter $b\sim 20$\,km\,s$^{-1}$, and
all vibrational transitions from the first vibrational mode, and the
$J=1-6$ transitions.  The templates were smoothed to match the
resolution of the composite spectra.  In fitting the models to the
composite spectra, the $\ebmv$, $\nhi$, and $f_{\rm cov}(\hi)$ were
fixed to the values found above, and $\nmolh$ and $f_{\rm cov}(\molh)$
were allowed to vary.  The ``best-fit'' $\nmolh$ and $f_{\rm
  cov}(\molh)$ were found by minimizing the differences in the model
and composite spectra in the four wavelength windows indicated by the
``$\nmolh$'' entry in Table~\ref{tab:waves}.  These windows were
chosen because they are free from $\hi$ absorption and strong metal
line absorption as identified in Far Ultraviolet Spectroscopic
Explorer (FUSE) spectra of the LMC by \citet{walborn02}.  The
best-fitting models including $\molh$ are shown in
Figure~\ref{fig:hih2fit}, and we find $f_{\rm cov}(\molh) \approx
0.15-0.19$ and $\lognmolh \approx 20.9$.  As the $\molh$ lines are
blended and unresolved, we cannot accurately determine the errors in
the fitted parameters.  Nevertheless, we note that the deficit in flux
of the composites relative to the models which {\em do not} include
the Lyman-Werner bands in the four $\molh$ windows
(Table~\ref{tab:waves}) implies that $\molh$ absorption must be
present, albeit with a low covering fraction or low column density
(Figure~\ref{fig:h2fit}).\footnote{The absorption arising from
  transitions between vibrational states above the ground vibrational
  state of $\molh$ is expected to be weak relative to the absorption
  arising from transitions from the ground vibrational state for the
  typical conditions in the ISM.  Transitions from the ground
  vibrational state (for the six rotational transitions considered
  here, $J=6-1$) completely blanket the far-UV continuum given the
  resolution of our spectra.  Thus, adding absorption corresponding to
  upper level vibrational transitions will simply remove ``power''
  from the considered (ground state vibrational) mode and the column
  densities and covering fractions derived here would be upper
  limits.}  The apparently low covering fraction of $\molh$ relative
to that of the $\hi$ and dust implies that most of the dust in the ISM
of typical galaxies at $z\sim 3$ is unrelated to the catalysis of
$\molh$ molecules, and is associated with other phases of the ISM
(e.g., ionized and neutral gas).  This result is robust against the
assumption of a non-unity covering fraction of dust.  In other words,
as shown in Reddy et~al. (2016b), assuming a {\em partial}
covering fraction of dust yields dust covering fractions that are $\ga
92\%$, still much larger than the covering fraction deduced for
$\molh$.

The aforementioned fitting demonstrates that the updated
\citet{reddy15} curve provides a reasonable fit to the composite
spectra over the entire far-UV through near-UV range once $\hi$ and
$\molh$ absorption are accounted for.  We found similar results when
considering the updated \citet{calzetti00} attenuation curve.
Moreover, the column density and covering fraction of $\molh$ indicate
a similar degree of absorption of the far-UV continuum between the
blue and red $\ebmv$ bins, implying that systematic differences in
this absorption are negligible and thus unlikely to bias our estimate
of the far-UV shape of the dust attenuation curve.  In comparison, the
original versions of these dust curves imply a far-UV rise that is too
steep to simultaneously model the mid-UV range used to compute $\ebmv$
($\lambda=1268-1633$\,\AA) and the far-UV at $\lambda \la 1050$\,\AA\,
(Figure~\ref{fig:hih2fit}).

\section{Discussion}
\label{sec:discussion}

A primary result of our analysis is that the far-UV attenuation curve
rises less rapidly toward bluer wavelengths ($\lambda \la 1250$\,\AA)
relative to other commonly assumed attenuation curves.  A consequence
of this less-rapid far-UV rise is a lower attenuation of LyC photons
by dust.  In particular, the average $\ebmv$ for galaxies in our
$z\sim 3$ sample is $\langle\ebmv\rangle_{\rm Calz} = 0.20$ and
$\langle\ebmv\rangle_{\rm MOS} = 0.21$ if we assume the
\citet{calzetti00} and \citet{reddy15} reddening curves, respectively.
For these values, the attenuation of LyC photons at
$\lambda=900$\,\AA\, is a factor of $\approx 1.8-2.3$ lower than that
inferred using polynomial extrapolations of the \citet{calzetti00} and
\citet{reddy15} curves at $\lambda<1200$\,\AA.  Clearly, this result
has implications for the fraction of hydrogen-ionizing photons that
escape galaxies before being absorbed by hydrogen in the IGM or CGM,
i.e., the ionizing escape fraction.  At face value, this result
implies a larger escape fraction of ionizing photons for galaxies of a
given dust reddening.  However, as discussed in Reddy et~al. (2016b),
the actual covering fraction of neutral gas likely plays a larger role
in modulating the escape of such photons than attenuation by dust.
Specifically, the dominant mode by which ionizing photons escape
$L^{\ast}$ galaxies at $z\sim 3$ is through channels that have been
cleared of gas and dust, while other sightlines are characterized by
optically-thick $\hi$ that extinguishes ionizing photons through
photoelectric absorption (Reddy et~al. 2016b).  Regardless, our new
determinations of the far-UV dust curve have direct applications to
assessing the attenuation of ionizing photons either internal to
\ion{H}{2} regions (e.g., \citealt{petrosian72, natta76, shields95,
  inoue01, dopita03}) or---in the context of high-redshift
galaxies---in the ionized galactic-scale ISM or in low column density
neutral gas (i.e., $\lognhi\la 17.2$), where dust absorption begins to
dominate the depletion of ionizing photons.

\begin{figure*}
\epsscale{1.00}
\plotone{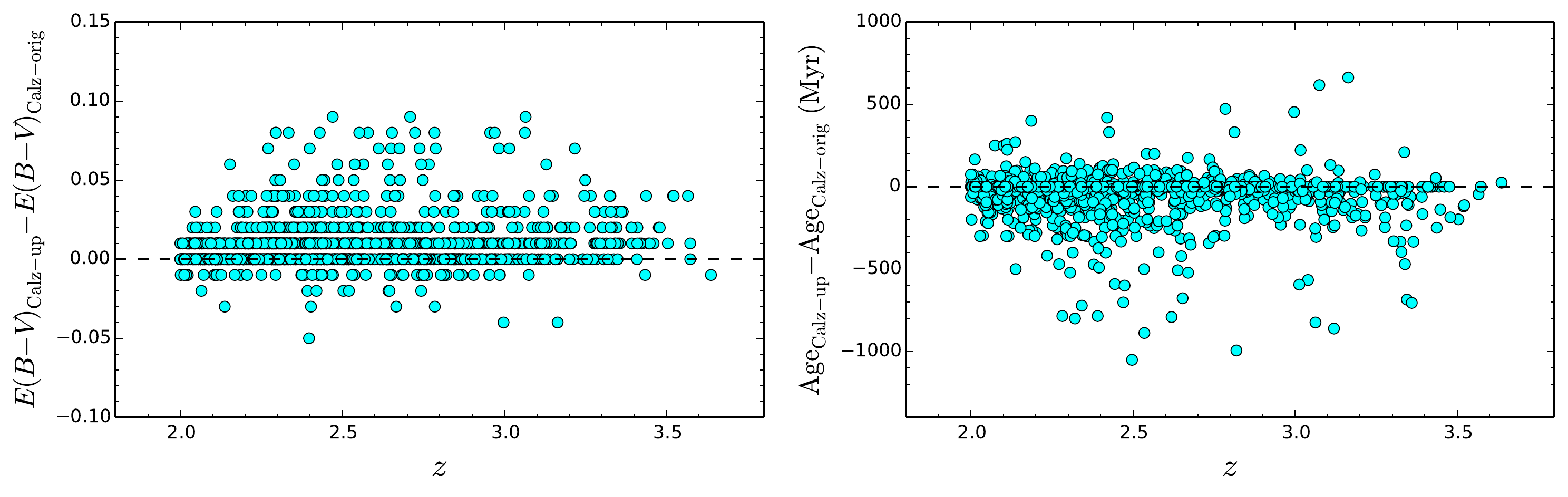}
\caption{Difference in the $\ebmv$ and ages derived from stellar
  population modeling for 1,874 KBSS galaxies when assuming the updated
  and original forms of the \citet{calzetti00} attenuation curve.  The
  updated form of the curve, which incorporates the new shape of the
  far-UV curve derived here, results in $\ebmv$ and ages that are on
  average $\approx 0.008$\,mag redder and $50$\,Myr younger,
  respectively, than those derived assuming the original
  \citet{calzetti00} curve.}
\label{fig:sedcomparison}
\end{figure*}

The revised shape of the far-UV curve also has consquences for the
reddening and ages of galaxies derived from stellar population (i.e.,
SED) modeling.  To illustrate this, we considered 1,874 galaxies from
KBSS that have ground-based $\ugr$, near-IR intermediate-band ($J2$,
$J3$, $H1$, $H2$), and broad-band ($J$, $\ks$) photometry, along with
{\em Hubble Space Telescope} WFC3/IR F160W and {\em Spitzer Space
  Telescope} IRAC imaging (Strom et~al., in prep).  We fit
\citet{bruzual03} $0.2Z_{\odot}$ stellar population models to the
photometry, assuming a constant star-formation history, $\ebmv$ in the
range 0.0-0.6, and ages from $50$\,Myr to the age of the Universe at
the redshift of each galaxy (a full description of the SED-fitting
procedure is provided in \citealt{reddy12b}).  The modeling was
performed assuming the original form of the \citet{calzetti00}
attenuation curve as well as the updated version (i.e., which
incorporates the new shape of the far-UV curve found here).
Figure~\ref{fig:sedcomparison} shows the comparison between the
reddening and ages derived for galaxies in KBSS as function of
redshift, for the two versions of the attenuation curve.  A less rapid
rise of the far-UV attenuation curve implies that a larger reddening
is required to reproduce the rest-UV color, which in turn makes the
galaxies appear younger.  Thus, the updated attenuation curve results
in $\ebmv$ that are on average $\approx 0.008$\,mag redder, and ages
that are on average $\approx 50$\,Myr younger, than those derived with
the original \citet{calzetti00} curve.  The redder $\ebmv$ result in
SFRs that are about $8\%$ larger, while the stellar masses do not
change appreciably.  Of course, the magnitude of the effect of
changing the far-UV attenuation curve on the derived stellar
population parameters depends on which bands are being fit.  For
galaxies at $z\sim 2-3$, where the Ly$\alpha$ forest is thin and the
$U_{\rm n}$ and $G$-bands probe the rest-frame far-UV, we find
systematic offsets in both the reddening and ages of galaxies.

\section{Summary}
\label{sec:summary}

To summarize, we have used a large sample of 933 spectroscopically
confirmed $\rs<25.5$ LBGs at $z\sim 3$, 121 of which have very deep
spectroscopic coverage in the range $850\la\lambda\la 1300$\,\AA, in
order to deduce the shape of the far-UV ($\lambda=950-1500$\,\AA) dust
attenuation curve.  We used an iterative procedure to construct
composite spectra of the $z\sim 3$ LBGs in bins of continuum color
excess, $\ebmv$, and, by taking ratios of these spectra, we calculated
the far-UV wavelength dependence of the dust attenuation curve.

The curve found here implies a lower attenuation at $\lambda \la
1250$\,\AA\, relative to that inferred from simple polynomial
extrapolations of the \citet{calzetti00} and \citet{reddy15}
attenuation curves.  Our new determination of the far-UV shape of the
dust curve enables more accurate estimates of the effect of dust on
the attenuation of hydrogen-ionizing (LyC) photons.  Specifically, our
results imply a factor of $\approx 2$ lower attenuation of LyC photons
for typical $\ebmv$ relative to that deduced from simple
extrapolations of the most widely used attenuation curves at high
redshift.  Our results may be used to assess the degree of LyC
attenuation internal to \ion{H}{2} regions, in the ionized media of
high-redshift galaxies, or in low column density ISM ($\lognhi \la
17.2$).

The composite spectra of $z\sim 3$ galaxies can be described well by a
model that assumes our new attenuation curve with a unity covering
fraction of dust, an $\approx 70-80\%$ covering fraction of $\hi$, and
additional absorption associated with the Lyman-Werner bands of
$\molh$.  The inferred covering fraction of $\molh$ is very low ($\la
20\%$) and, when contrasted with the high covering fraction of dust,
suggests that most of the dust in the ISM of high-redshift galaxies is
not associated with the catalysis of $\molh$.  The dust is likely
associated with other phases of the ISM such as the ionized and/or
neutral $\hi$ components. 

\acknowledgements

NAR is supported by an Alfred P. Sloan Research Fellowship, and
acknowledges the visitors program at the Institute of Astronomy in
Cambridge, UK, where part of this research was conducted.  CCS
acknowledges NSF grants AST-0908805 and AST-1313472.  MB acknowledges
support of the Serbian MESTD through grant ON176021.  We thank Allison
Strom for providing the photometry for the KBSS galaxies.  We are
grateful to the anonymous referee whose comments led to significant
improvements in the presentation of the analysis.  We wish to extend
special thanks to those of Hawaiian ancestry on whose sacred mountain
we are privileged to be guests.  Without their generous hospitality,
the observations presented herein would not have been possible.


\begin{thebibliography}{57}
\expandafter\ifx\csname natexlab\endcsname\relax\def\natexlab#1{#1}\fi

\bibitem[{{Adelberger} {et~al.}(2004){Adelberger}, {Steidel}, {Shapley},
  {Hunt}, {Erb}, {Reddy}, \& {Pettini}}]{adelberger04}
{Adelberger}, K.~L., {Steidel}, C.~C., {Shapley}, A.~E., et~al. 2004, \apj, 607, 226

\bibitem[{{Asplund} {et~al.}(2009){Asplund}, {Grevesse}, {Sauval}, \&
  {Scott}}]{asplund09}
{Asplund}, M., {Grevesse}, N., {Sauval}, A.~J., \& {Scott}, P. 2009, \araa, 47,
  481

\bibitem[{{Atek} {et~al.}(2014){Atek}, {Kunth}, {Schaerer}, {Mas-Hesse},
  {Hayes}, {{\"O}stlin}, \& {Kneib}}]{atek14}
{Atek}, H., {Kunth}, D., {Schaerer}, D., et~al. 2014, \aap, 561, A89

\bibitem[{{Brott} {et~al.}(2011){Brott}, {de Mink}, {Cantiello}, {Langer}, {de
  Koter}, {Evans}, {Hunter}, {Trundle}, \& {Vink}}]{brott11}
{Brott}, I., {de Mink}, S.~E., {Cantiello}, M., et~al. 2011, \aap,
  530, A115

\bibitem[{{Bruzual} \& {Charlot}(2003)}]{bruzual03}
{Bruzual}, G. \& {Charlot}, S. 2003, \mnras, 344, 1000

\bibitem[{{Buat} {et~al.}(2011){Buat}, {Giovannoli}, {Heinis}, {Charmandaris},
  {Coia}, {Daddi}, {Dickinson}, {Elbaz}, {Hwang}, {Morrison}, {Dasyra},
  {Aussel}, {Altieri}, {Dannerbauer}, {Kartaltepe}, {Leiton}, {Magdis},
  {Magnelli}, \& {Popesso}}]{buat11}
{Buat}, V., {Giovannoli}, E., {Heinis}, S., et~al. 2011, \aap,
  533, A93

\bibitem[{{Buat} {et~al.}(2012){Buat}, {Noll}, {Burgarella}, {Giovannoli},
  {Charmandaris}, {Pannella}, {Hwang}, {Elbaz}, {Dickinson}, {Magdis}, {Reddy},
  \& {Murphy}}]{buat12}
{Buat}, V., {Noll}, S., {Burgarella}, D., et~al. 2012, \aap, 545, A141

\bibitem[{{Calzetti} {et~al.}(2000){Calzetti}, {Armus}, {Bohlin}, {Kinney},
  {Koornneef}, \& {Storchi-Bergmann}}]{calzetti00}
{Calzetti}, D., {Armus}, L., {Bohlin}, R.~C., et~al. 2000, \apj, 533, 682

\bibitem[{{Calzetti} \& {Kinney}(1992)}]{calzetti92}
{Calzetti}, D. \& {Kinney}, A.~L. 1992, \apjl, 399, L39

\bibitem[{{Calzetti} {et~al.}(1994){Calzetti}, {Kinney}, \&
  {Storchi-Bergmann}}]{calzetti94}
{Calzetti}, D., {Kinney}, A.~L., \& {Storchi-Bergmann}, T. 1994, \apj, 429, 582

\bibitem[{{Charlot} \& {Fall}(1993)}]{charlot93}
{Charlot}, S. \& {Fall}, S.~M. 1993, \apj, 415, 580

\bibitem[{{Dopita} {et~al.}(2003){Dopita}, {Groves}, {Sutherland}, \&
  {Kewley}}]{dopita03}
{Dopita}, M.~A., {Groves}, B.~A., {Sutherland}, R.~S., \& {Kewley}, L.~J. 2003,
  \apj, 583, 727

\bibitem[{{Eldridge} \& {Stanway}(2009)}]{eldridge09}
{Eldridge}, J.~J. \& {Stanway}, E.~R. 2009, \mnras, 400, 1019

\bibitem[{{Finkelstein} {et~al.}(2011){Finkelstein}, {Cohen}, {Moustakas},
  {Malhotra}, {Rhoads}, \& {Papovich}}]{finkelstein11}
{Finkelstein}, S.~L., {Cohen}, S.~H., {Moustakas}, J., et~al. 2011, \apj, 733, 117

\bibitem[{{Giavalisco} {et~al.}(1996){Giavalisco}, {Koratkar}, \&
  {Calzetti}}]{giavalisco96}
{Giavalisco}, M., {Koratkar}, A., \& {Calzetti}, D. 1996, \apj, 466, 831

\bibitem[{{Gordon} {et~al.}(2009){Gordon}, {Cartledge}, \&
  {Clayton}}]{gordon09}
{Gordon}, K.~D., {Cartledge}, S., \& {Clayton}, G.~C. 2009, \apj, 705, 1320

\bibitem[{{Hayes} {et~al.}(2011){Hayes}, {Schaerer}, {{\"O}stlin}, {Mas-Hesse},
  {Atek}, \& {Kunth}}]{hayes11}
{Hayes}, M., {Schaerer}, D., {{\"O}stlin}, G., et~al. 2011, \apj, 730, 8

\bibitem[{{Inoue}(2001)}]{inoue01}
{Inoue}, A.~K. 2001, \aj, 122, 1788

\bibitem[{{Johnson} {et~al.}(2007){Johnson}, {Schiminovich}, {Seibert},
  {Treyer}, {Martin}, {Barlow}, {Forster}, {Friedman}, {Morrissey}, {Neff},
  {Small}, {Wyder}, {Bianchi}, {Donas}, {Heckman}, {Lee}, {Madore}, {Milliard},
  {Rich}, {Szalay}, {Welsh}, \& {Yi}}]{johnson07a}
{Johnson}, B.~D., {Schiminovich}, D., {Seibert}, M., et~al. 2007, \apjs, 173, 377

\bibitem[{{Kriek} \& {Conroy}(2013)}]{kriek13}
{Kriek}, M. \& {Conroy}, C. 2013, \apjl, 775, L16

\bibitem[{{Kriek} {et~al.}(2015){Kriek}, {Shapley}, {Reddy}, {Siana}, {Coil},
  {Mobasher}, {Freeman}, {de Groot}, {Price}, {Sanders}, {Shivaei}, {Brammer},
  {Momcheva}, {Skelton}, {van Dokkum}, {Whitaker}, {Aird}, {Azadi}, {Kassis},
  {Bullock}, {Conroy}, {Dav{\'e}}, {Kere{\v s}}, \& {Krumholz}}]{kriek15}
{Kriek}, M., {Shapley}, A.~E., {Reddy}, N.~A., et~al. 2015, \apjs, 218, 15

\bibitem[{{Leitherer} {et~al.}(2014){Leitherer}, {Ekstr{\"o}m}, {Meynet},
  {Schaerer}, {Agienko}, \& {Levesque}}]{leitherer14}
{Leitherer}, C., {Ekstr{\"o}m}, S., {Meynet}, G., et~al. 2014, \apjs, 212, 14

\bibitem[{{Leitherer} {et~al.}(2002){Leitherer}, {Li}, {Calzetti}, \&
  {Heckman}}]{leitherer02}
{Leitherer}, C., {Li}, I.-H., {Calzetti}, D., \& {Heckman}, T.~M. 2002, \apjs,
  140, 303

\bibitem[{{Leitherer} {et~al.}(2010){Leitherer}, {Ortiz Ot{\'a}lvaro},
  {Bresolin}, {Kudritzki}, {Lo Faro}, {Pauldrach}, {Pettini}, \&
  {Rix}}]{leitherer10}
{Leitherer}, C., {Ortiz Ot{\'a}lvaro}, P.~A., {Bresolin}, F., et~al. 2010, \apjs, 189, 309

\bibitem[{{Leitherer} {et~al.}(1999){Leitherer}, {Schaerer}, {Goldader},
  {Delgado}, {Robert}, {Kune}, {de Mello}, {Devost}, \&
  {Heckman}}]{leitherer99}
{Leitherer}, C., {Schaerer}, D., {Goldader}, J.~D., et~al. 1999, \apjs, 123, 3

\bibitem[{{Levesque} {et~al.}(2012){Levesque}, {Leitherer}, {Ekstrom},
  {Meynet}, \& {Schaerer}}]{levesque12}
{Levesque}, E.~M., {Leitherer}, C., {Ekstrom}, S., {Meynet}, G., \& {Schaerer},
  D. 2012, \apj, 751, 67

\bibitem[{{McCandliss}(2003)}]{mccandliss03}
{McCandliss}, S.~R. 2003, \pasp, 115, 651

\bibitem[{{Meier} \& {Terlevich}(1981)}]{meier81}
{Meier}, D.~L. \& {Terlevich}, R. 1981, \apjl, 246, L109

\bibitem[{{Meurer} {et~al.}(1999){Meurer}, {Heckman}, \& {Calzetti}}]{meurer99}
{Meurer}, G.~R., {Heckman}, T.~M., \& {Calzetti}, D. 1999, \apj, 521, 64

\bibitem[{{Natta} \& {Panagia}(1976)}]{natta76}
{Natta}, A. \& {Panagia}, N. 1976, \aap, 50, 191

\bibitem[{{Noll} {et~al.}(2009){Noll}, {Pierini}, {Cimatti}, {Daddi}, {Kurk},
  {Bolzonella}, {Cassata}, {Halliday}, {Mignoli}, {Pozzetti}, {Renzini},
  {Berta}, {Dickinson}, {Franceschini}, {Rodighiero}, {Rosati}, \&
  {Zamorani}}]{noll09}
{Noll}, S., {Pierini}, D., {Cimatti}, A., et~al. 2009, \aap, 499, 69

\bibitem[{{Oke} {et~al.}(1995){Oke}, {Cohen}, {Carr}, {Cromer}, {Dingizian},
  {Harris}, {Labrecque}, {Lucinio}, {Schaal}, {Epps}, \& {Miller}}]{oke95}
{Oke}, J.~B., {Cohen}, J.~G., {Carr}, M., et~al. 1995, \pasp, 107, 375

\bibitem[{{Petrosian} {et~al.}(1972){Petrosian}, {Silk}, \&
  {Field}}]{petrosian72}
{Petrosian}, V., {Silk}, J., \& {Field}, G.~B. 1972, \apjl, 177, L69

\bibitem[{{Reddy} {et~al.}(2015){Reddy}, {Kriek}, {Shapley}, {Freeman},
  {Siana}, {Coil}, {Mobasher}, {Price}, {Sanders}, \& {Shivaei}}]{reddy15}
{Reddy}, N.~A., {Kriek}, M., {Shapley}, A.~E., et~al. 2015, \apj, 806, 259

\bibitem[{{Reddy} {et~al.}(2012){Reddy}, {Pettini}, {Steidel}, {Shapley},
  {Erb}, \& {Law}}]{reddy12b}
{Reddy}, N.~A., {Pettini}, M., {Steidel}, C.~C., et~al. 2012, \apj, 754, 25

\bibitem[{{Reddy} {et~al.}(2006){Reddy}, {Steidel}, {Erb}, {Shapley}, \&
  {Pettini}}]{reddy06b}
{Reddy}, N.~A., {Steidel}, C.~C., {Erb}, D.~K., {Shapley}, A.~E., \& {Pettini},
  M. 2006, \apj, 653, 1004

\bibitem[{{Reddy} {et~al.}(2008){Reddy}, {Steidel}, {Pettini}, {Adelberger},
  {Shapley}, {Erb}, \& {Dickinson}}]{reddy08}
{Reddy}, N.~A., {Steidel}, C.~C., {Pettini}, M., et~al. 2008, \apjs, 175, 48

\bibitem[{{Rix} {et~al.}(2004){Rix}, {Pettini}, {Leitherer}, {Bresolin},
  {Kudritzki}, \& {Steidel}}]{rix04}
{Rix}, S.~A., {Pettini}, M., {Leitherer}, C., et~al. 2004, \apj, 615, 98

\bibitem[{{Rudie} {et~al.}(2013){Rudie}, {Steidel}, {Shapley}, \&
  {Pettini}}]{rudie13}
{Rudie}, G.~C., {Steidel}, C.~C., {Shapley}, A.~E., \& {Pettini}, M. 2013,
  \apj, 769, 146

\bibitem[{{Salmon} {et~al.}(2015){Salmon}, {Papovich}, {Long}, {Willner},
  {Finkelstein}, {Ferguson}, {Dickinson}, {Duncan}, {Faber}, {Hathi},
  {Koekemoer}, {Kurczynski}, {Newman}, {Pacifici}, {Perez-Gonzalez}, \&
  {Pforr}}]{salmon15}
{Salmon}, B., {Papovich}, C., {Long}, J., et~al. 2015, ArXiv e-prints

\bibitem[{{Sawicki} \& {Thompson}(2006)}]{sawicki06}
{Sawicki}, M. \& {Thompson}, D. 2006, \apj, 642, 653

\bibitem[{{Scarlata} {et~al.}(2009){Scarlata}, {Colbert}, {Teplitz}, {Panagia},
  {Hayes}, {Siana}, {Rau}, {Francis}, {Caon}, {Pizzella}, \&
  {Bridge}}]{scarlata09}
{Scarlata}, C., {Colbert}, J., {Teplitz}, H.~I., et~al. 2009, \apjl, 704, L98

\bibitem[{{Scoville} {et~al.}(2015){Scoville}, {Faisst}, {Capak}, {Kakazu},
  {Li}, \& {Steinhardt}}]{scoville15}
{Scoville}, N., {Faisst}, A., {Capak}, P., et~al. 2015, \apj, 800, 108

\bibitem[{{Shapley} {et~al.}(2015){Shapley}, {Reddy}, {Kriek}, {Freeman},
  {Sanders}, {Siana}, {Coil}, {Mobasher}, {Shivaei}, {Price}, \& {de
  Groot}}]{shapley15}
{Shapley}, A.~E., {Reddy}, N.~A., {Kriek}, M., et~al. 2015, \apj, 801, 88

\bibitem[{{Shapley} {et~al.}(2001){Shapley}, {Steidel}, {Adelberger},
  {Dickinson}, {Giavalisco}, \& {Pettini}}]{shapley01}
{Shapley}, A.~E., {Steidel}, C.~C., {Adelberger}, K.~L., {Dickinson}, M.,
  {Giavalisco}, M., \& {Pettini}, M. 2001, \apj, 562, 95

\bibitem[{{Shields} \& {Kennicutt}(1995)}]{shields95}
{Shields}, J.~C. \& {Kennicutt}, Jr., R.~C. 1995, \apj, 454, 807

\bibitem[{{Steidel} {et~al.}(2003){Steidel}, {Adelberger}, {Shapley},
  {Pettini}, {Dickinson}, \& {Giavalisco}}]{steidel03}
{Steidel}, C.~C., {Adelberger}, K.~L., {Shapley}, A.~E., et~al. 2003, \apj, 592, 728

\bibitem[{{Steidel} {et~al.}(2010){Steidel}, {Erb}, {Shapley}, {Pettini},
  {Reddy}, {Bogosavljevi{\'c}}, {Rudie}, \& {Rakic}}]{steidel10}
{Steidel}, C.~C., {Erb}, D.~K., {Shapley}, A.~E., et~al. 2010, \apj, 717, 289

\bibitem[{{Steidel} \& {Hamilton}(1993)}]{steidel93}
{Steidel}, C.~C. \& {Hamilton}, D. 1993, \aj, 105, 2017

\bibitem[{{Steidel} {et~al.}(2014){Steidel}, {Rudie}, {Strom}, {Pettini},
  {Reddy}, {Shapley}, {Trainor}, {Erb}, {Turner}, {Konidaris}, {Kulas}, {Mace},
  {Matthews}, \& {McLean}}]{steidel14}
{Steidel}, C.~C., {Rudie}, G.~C., {Strom}, A.~L., et~al. 2014, \apj, 795, 165

\bibitem[{{Steidel} {et~al.}(2004){Steidel}, {Shapley}, {Pettini},
  {Adelberger}, {Erb}, {Reddy}, \& {Hunt}}]{steidel04}
{Steidel}, C.~C., {Shapley}, A.~E., {Pettini}, M., et~al. 2004, \apj, 604, 534

\bibitem[{{Troncoso} {et~al.}(2014){Troncoso}, {Maiolino}, {Sommariva},
  {Cresci}, {Mannucci}, {Marconi}, {Meneghetti}, {Grazian}, {Cimatti},
  {Fontana}, {Nagao}, \& {Pentericci}}]{troncoso14}
{Troncoso}, P., {Maiolino}, R., {Sommariva}, V., et~al. 2014, \aap, 563, A58

\bibitem[{{Verhamme} {et~al.}(2008){Verhamme}, {Schaerer}, {Atek}, \&
  {Tapken}}]{verhamme08}
{Verhamme}, A., {Schaerer}, D., {Atek}, H., \& {Tapken}, C. 2008, \aap, 491, 89

\bibitem[{{Walborn} {et~al.}(2002){Walborn}, {Fullerton}, {Crowther},
  {Bianchi}, {Hutchings}, {Pellerin}, {Sonneborn}, \& {Willis}}]{walborn02}
{Walborn}, N.~R., {Fullerton}, A.~W., {Crowther}, P.~A., et~al. 2002,
  \apjs, 141, 443

\bibitem[{{Weingartner} \& {Draine}(2001)}]{weingartner01}
{Weingartner}, J.~C. \& {Draine}, B.~T. 2001, \apj, 548, 296

\bibitem[{{Wild} {et~al.}(2011){Wild}, {Charlot}, {Brinchmann}, {Heckman},
  {Vince}, {Pacifici}, \& {Chevallard}}]{wild11}
{Wild}, V., {Charlot}, S., {Brinchmann}, J., et~al. 2011, \mnras, 417, 1760

\bibitem[{{Zeimann} {et~al.}(2015){Zeimann}, {Ciardullo}, {Gronwall}, {Bridge},
  {Brooks}, {Fox}, {Gawiser}, {Gebhardt}, {Hagen}, {Schneider}, \&
  {Trump}}]{zeimann15}
{Zeimann}, G.~R., {Ciardullo}, R., {Gronwall}, C., et~al. 2015, \apj, 814, 162

\end{thebibliography}








\end{document}